\documentclass[namedreferences]{SolarPhysics}
\usepackage[optionalrh]{spr-sola-addons} 
\usepackage{graphicx}                    
\usepackage{color}                       
\usepackage{url}                         

\newcommand{\etal}{{\it et al.}}
\newcommand{\eg}{{\it e.g.}}

\begin{document}

\begin{article}

\begin{opening}

\title{ROSA: a high cadence, synchronized multi-camera solar imaging system}

%
\author{D.~B.~\surname{Jess}$^{1}$\sep
        M.~\surname{Mathioudakis}$^{1}$\sep
        D.~J.~\surname{Christian}$^{2}$\sep
	F.~P.~\surname{Keenan}$^{1}$\sep
	R.~S.~I.~\surname{Ryans}$^{1}$\sep
	P.~J.~\surname{Crockett}$^{1}$
	}

%
\runningauthor{D.~B.~Jess et al.}
\runningtitle{A ROSA view of the Sun}

%
  \institute{$^{1}$ Astrophysics Research Centre, School of Mathematics and Physics, Queen's University Belfast, Belfast, BT7~1NN, Northern Ireland, U.K.
                     email: \url{d.jess@qub.ac.uk} \\ 
             $^{2}$ Department of Physics and Astronomy, California State University Northridge, 18111 Nordhoff Street, Northridge, CA 91330, USA.\\
             }

\begin{abstract}
Rapid Oscillations in the Solar Atmosphere (ROSA) is a synchronized, six camera high cadence solar imaging 
instrument developed by Queen's University Belfast. The system is available on the Dunn Solar Telescope 
at the National Solar Observatory in Sunspot, New Mexico, USA as a common-user instrument. Consisting of 
six 1k~$\times$~1k Peltier-cooled frame-transfer CCD cameras with very low noise (0.02 -- 15 e~s$^{-1}$~pixel$^{-1}$), 
each ROSA camera is capable of full-chip readout speeds in excess of 30~Hz, or 200~Hz when the CCD is windowed. 
Combining multiple cameras and fast readout rates, ROSA will accumulate approximately 12~TB of data per 8~hours 
observing. Following successful commissioning during August 2008, ROSA will 
allow multi-wavelength studies of the solar atmosphere at high temporal resolution. 
\end{abstract}

%

\end{opening}

%
\section{Introduction}
\label{introduction} 

High cadence observations of astronomical sources is a growing field within astrophysics, and there is a 
clear need for such data for the Sun. Many research topics, in particular those related to the 
dynamic Sun and the heating of its outer regions, involve the observation and modelling of wave phenomena and 
explosive events captured over very short timescales. High cadence observations are also important for post-facto 
image reconstruction (PFIR) techniques which require the processing of an extensive collection of images for the 
production of a single frame at diffraction-limited resolution. These short exposure images must be accumulated over 
timescales sufficiently small so that atmospheric turbulence is effectively ``frozen out'', and the solar features 
remain unchanged.

Since the discovery of solar oscillations in the 1960s \cite{Lei60}, and their subsequent 
confirmation by \inlinecite{Deu75}, there has been a multitude of observational evidence 
presented verifying the widespread existence of oscillations in the solar atmosphere (see the recent review 
by \opencite{Ban07}). Oscillations throughout the solar atmosphere have enabled scientists to 
probe the underlying physics that are involved with energy transfer and coronal heating, and 
have been suggested as candidates to explain one of the main unanswered questions in solar 
physics --  {\it Why is the outer solar atmosphere hotter than its surface?} However, in recent 
years a need to probe high-frequency oscillations has arisen. \inlinecite{Por94} have shown that 
the energy contribution of high-frequency waves to atmospheric heating may be significant. 
Furthermore, \inlinecite{Has08} have indicated that quasi-periodic oscillations 
shorter than 100~s may be a vital mechanism in the creation of the multi-million degree solar 
corona. Thus, a sensitive camera system with a suitably high cadence to satisfy the 
Nyquist parameter is paramount for the study of high-frequency oscillations. 

Such a system is not only capable of studying wave motion, but may also be utilized to 
examine highly dynamic phenomena in the lower solar atmosphere. Recently, fast-moving plasma 
with velocities exceeding 200~kms$^{-1}$ has been observed by \inlinecite{Van06}, indicating the abundance 
of highly dynamic structures in the chromosphere. Additional investigations may include the search for hard 
X-ray non-thermal electron precipitation sites in the event of flare activity.  \inlinecite{Kip83} report 
short duration spikes associated with hard X-ray emission from solar flares, and in conjunction with the 
impact of fast-moving non-thermal electrons, leads to enhanced emission over inherently short time scales 
($<~0.04$~s; \opencite{Bro71}). This, coupled with the detection of fast fluctuations of H$\alpha$ emission 
from a flare kernel on timescales of $0.3 - 0.7$~s by \inlinecite{Wan00}, demonstrates the need for high-cadence 
solar imaging techniques to capture these dynamic processes on their intrinsic scales.

The search for rapid, often low-amplitude intensity variations requires a highly sensitive camera system 
providing accurate, sustained frame rates, accompanied by good local seeing at the telescope facility. 
Here we report the construction of a new high-speed camera system, developed by Queen's University Belfast (QUB), 
and its commissioning as a common-user instrument at the Dunn Solar Telescope (DST) facility run by the National 
Solar Observatory in the Sacramento Peak mountains, New Mexico, USA. This system, named 
{\em{Rapid Oscillations in the Solar Atmosphere}} (ROSA), will present users with the ability to observe the 
Sun simultaneously in up to six wavelengths or wavebands. A detailed guide to ROSA's hardware is 
described in Section~\ref{instrumentation}, while in Section~\ref{observing} a description of available observing 
modes is presented. In Section~\ref{interface} ROSA's graphical user interface (GUI) is explained, 
followed by guidelines for obtaining calibrated data in Section~\ref{calibration}. Finally, 
concluding remarks, in addition to sample images from the instrument commissioning, are given in 
Section~\ref{conclusions}.

\section{Instrumentation}
\label{instrumentation}

In the late 1990's, a high-speed, two-camera solar imaging system named the {\em{Solar Eclipse Corona Imaging 
System}} (SECIS; \opencite{Phi00}) began development. This instrument was designed to rapidly capture 
images of the solar corona during a total eclipse and produced some very interesting results 
(see {\eg} \opencite{Wil01}, \citeyear{Wil02}; \opencite{Kat03}). However, in order to image 
the lower solar atmosphere at high time resolution, a new camera system needed to be developed. The {\em{Rapid 
Dual Imager}} (RDI; \opencite{Jes07a}) was designed as a followup to the highly successful SECIS camera 
system and consists of two 502~$\times$~494 pixel$^2$ CCDs, operating at speeds of up to 20 frames per 
second. Even with a relatively small number of pixels and onboard storage, RDI was able 
to show the need for a dedicated high-cadence multi-camera system (see {\eg} \opencite{Jes07b}) and 
acted as a pre-cursor to the capabilities possessed by ROSA.

The ROSA system consists of six individual frame transfer CCDs, each with their own dedicated server for data 
acquisition and storage. All cameras are triggered via a Precision Control Unit or ``sync'' box. 
The CCD cameras are from Andor Technologies of Belfast, with the DU-885-VP model 
chosen comprising of a $1004~\times~1002$ pixel$^{2}$ area. They are cooled up to 100~$^{\circ}$C 
below ambient temperature through use of a thermoelectric Peltier cooler. Maintaining low readout 
noise necessary for fast readout rates, each ROSA camera can read out over 30 frames per second in 
full-chip mode and over 200 frames per second when the CCD is windowed. The CCD characteristics 
specific to the ROSA system are presented in Table~\ref{table1} and further described in 
Section~\ref{EMCCD}.

\begin{table}
\caption{Detailed description of ROSA camera characteristics.}
\label{table1}
\begin{tabular}{lc}     
  \hline                   
Camera Parameter & Specification \\
  \hline
Model                         		& Andor iXon+ DU-885K-VP \\
Pixels                        		& 1004 $\times$ 1002 \\
Pixel Size                    		& $8 \times 8 \mu$m \\
Maximum Frame rate (full chip)        	& $30$ s$^{-1}$ \\
Maximum Frame rate (windowed chip)    	& $200$ s$^{-1}$ \\
Read noise                    		& 0.02 -- 15 e~s$^{-1}$~pixel$^{-1}$ \\
Data output                   		& 14 bit \\
\hline
\end{tabular}
\end{table}

The sync box provides independent 5~V DC pulses to each of the ROSA cameras to control the frame 
acquisition at preset trigger rates. Four trigger rates can be specified by the telescope user 
based upon the filters and exposure times chosen. As a result, all six cameras can be triggered 
at the same synchronized frame rate, or by any combination of the user-defined trigger rates. The 
sync box receives information from the master computer through a USB connection and is hard-wired 
to each ROSA camera via individual 5~m cables. Timing errors related to the arrival of 
synchronous pulses from different trigger cables are $<30\mu$s. 

\begin{figure} 
\centerline{\includegraphics[width=1.0\textwidth,clip=]{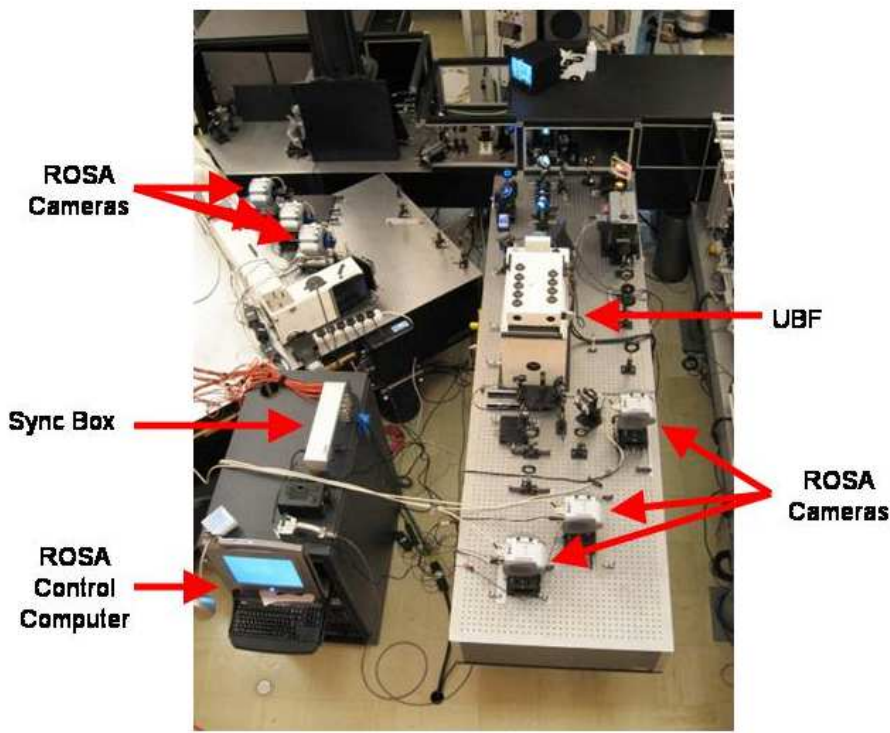}}
\caption{Six ROSA cameras placed on two optical benches at the DST during the commissioning run in August 2008. Three cameras 
are positioned behind ``blue'' filters (G-band, 4170~{\AA} continuum and Ca-K core; left optical bench), while 
the remaining three cameras are positioned behind ``red'' filters (H$\alpha$ core and circularly polarized measurements 
at 6302~{\AA}; right optical bench). One of the rack-mounted storage assemblies is also visible towards the lower-left side 
of the image.}
\label{ROSA_bench}
\end{figure}

Each camera has its own dedicated high-speed server that controls the data acquisition and storage. 
The PCs are Dell PowerEdge 2900 dual-core Xeon units, each with 4~GB RAM and over 1~TB of onboard 
storage. To achieve sustained frame rates and prevent data corruption, each ROSA server consists of 
eight high-speed (15000~RPM) hard drives, running in a RAID~0 configuration to boost available 
storage. All six servers are mounted in two, wheeled-rack enclosures, allowing the entire system to 
be re-positioned easily. Control of the six independent PCs can be achieved through use of a master 
server (also stored within the rack-mounted enclosure), where a KVM module allows each individual 
computer's display to be piped to a single screen.

Running at a sustained frame rate of 30~frames per second in all six ROSA cameras, approximately 
1.3~TB of data are accumulated every hour. The data are written into flexible image transport 
system (FITS; \opencite{Han01}) format incorporating a header and multiple image extensions. 
Detailed information related to the observing sequence are written in the main FITS header, and 
include descriptions of both the CCD and observing parameters, in addition to the acquisition start 
time. Each FITS file contains 256 individual images to keep the file size manageable, with each 
image written to a separate extension including only a time stamp in its header information.  There 
are currently two available options to transfer ROSA data to external media. The first is through use of three LTO3 tape 
autoloaders attached to the ROSA instrument. 
Each tape autoloader can hold ten 400~GB uncompressed LTO3 tapes, providing the ability to backup 
approximately 11~TB of data. An alternative mechanism to transport data is via user-supplied external 
media. Hard drives equipped with a USB adaptor can connect directly to the storage computers, thus 
allowing the transfer of data to external media quickly and avoiding the bottleneck created when 
using traditional FTP or SSH commands over a network. It is anticipated that an e-SATA connection 
will be made available on each ROSA storage computer in the near future to further accelerate data backups.

\subsection{Electron Multiplying CCDs}
\label{EMCCD}

A solar imaging system capable of acquiring high sustained frame rates places unprecedented 
demands on detector technology. In particular, the short exposure times needed to maintain 
high cadence imaging often leads to photon starvation, producing grainy and lacklustre solar 
images. However, in early 2000, Electron Multiplying Charge Coupled Devices 
(EMCCDs) were developed as image sensors capable of detecting single photon events without an image 
intensifier. This was achieved by way of a unique electron multiplying 
structure built directly into the chip and enables solar observers to attain high-contrast images without the 
drawbacks associated with long exposure times \cite{Den03}. 

\begin{figure} 
\centerline{\includegraphics[width=0.9\textwidth,clip=]{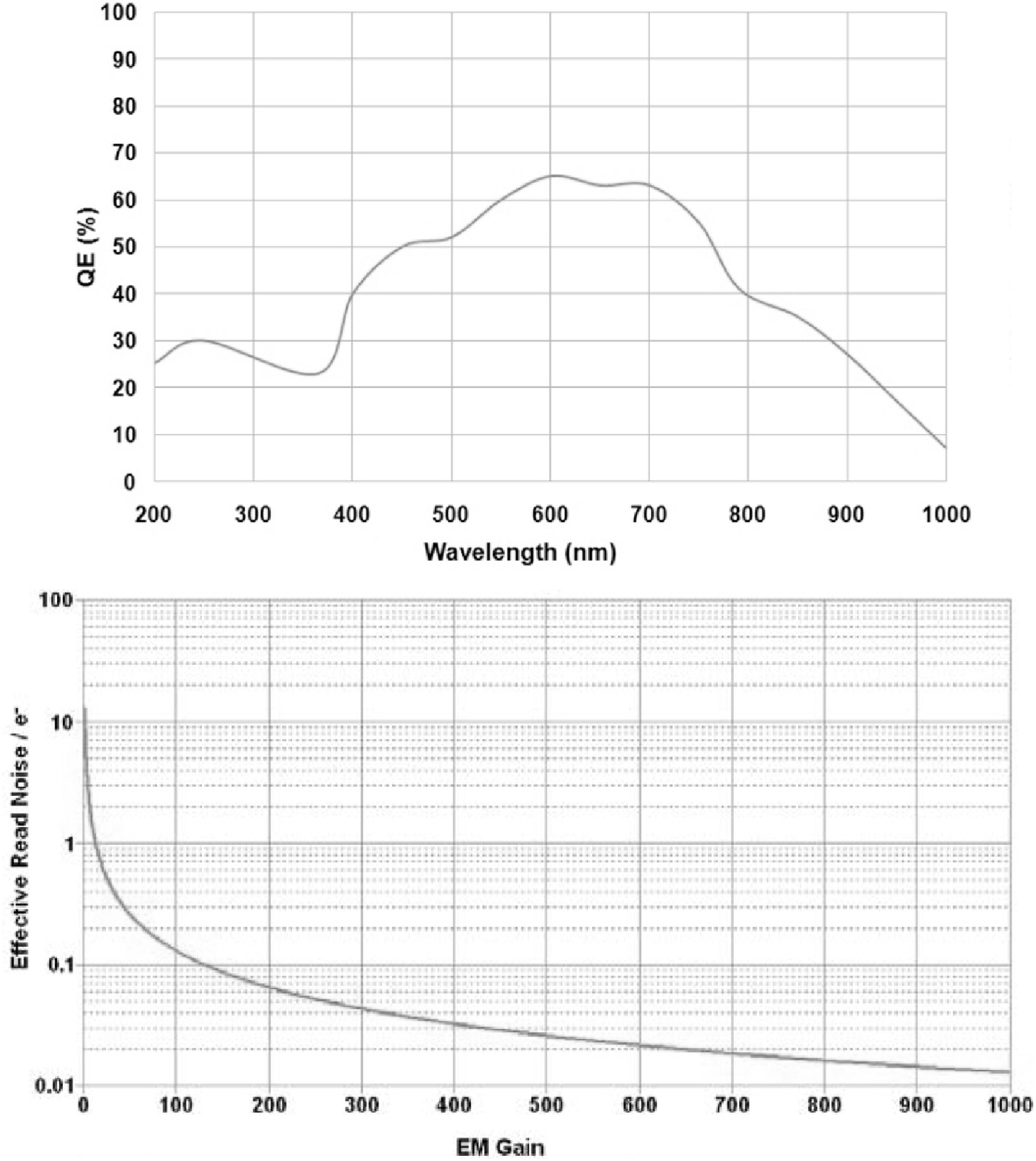}}
\caption{Upper panel is a representation of a ROSA camera's quantum efficiency as a function of wavelength, while the bottom panel 
plots a ROSA camera's effective read noise as a function of electron-multiplying gain. In this instance, the effective read noise is a 
combination of CCD readout noise and analogue-to-digital convertor noise.}
\label{QE_Noise}
\end{figure}

A key characteristic of EMCCDs is that they are not limited by the readout noise of the chip's 
output amplifier like a conventional CCD, even when operated at high readout speeds. This is 
achieved by allowing weak signals 
to be multiplied before any readout noise is added through the output amplifier through the addition of a solid-state 
electron multiplying register to the end of the normal serial register \cite{Coa04}. 
Camera readout noise is approximately 15~e~s$^{-1}$~pixel$^{-1}$. This measurement is for the entire system, and 
includes a combination of CCD readout noise and analogue-to-digital convertor noise. 
The value is for single-pixel readout with a zero second exposure under dark conditions with unit electron-multiplication gain. 
The electron multiplying register 
has several hundred stages that use higher than normal clock voltages allowing impact ionization of 
secondary electrons, and hence electron gain. Through use of hundreds of stages, the 
resultant gain can be user controlled to provide amplification ranging from unity to more than several thousand. 
In this regime, the effective system readout noise is reduced to below 1~e~s$^{-1}$~pixel$^{-1}$, often reaching 
0.02~e~s$^{-1}$~pixel$^{-1}$ under optimal conditions. Typical variations of readout noise with respect to electron multiplication 
is displayed in the lower panel of Figure~\ref{QE_Noise}.

Due to EMCCDs not requiring an image intensifier, it means that the full quantum efficiency of a silicon 
sensor, which can be as high as 95\%, may be utilized. 
The quantum efficiencies of the ROSA cameras with respect to incident wavelength is plotted in the 
upper panel of Figure~\ref{QE_Noise}. 
It is the combination of a minimal noise floor and a high 
quantum efficiency which renders the new breed of EMCCDs the most suitable for high-cadence solar imaging systems. 
However, it must be noted that the electron multiplication process will inherently introduce additional noise. As 
a result, electron multiplication is best implemented when short exposure times and low-transmission filters result in weak incident 
photon counts ({\it{i.e.}} when CCD readout noise dominates). Furthermore, the application of gain will reduce the number of 
camera applications by bringing the saturation point to lower flux levels. Care must therefore be taken when setting 
an electron-multiplication gain that maximizes the detection of faint features while still keeping brighter areas below their 
saturation point.
Physically, EMCCDs resemble a conventional frame transfer CCD structure, whereby an image is captured in the exposed 
region, before being shifted behind a masked storage area and read out.

\begin{table}
\caption{Parameters related to key optical filters which may be used in conjunction with ROSA. Values which still require 
clarification are listed as ``TBD''.}
\label{table2}
\begin{tabular}{lccccc}     
\hline
Filter Name      &  Central     &  Filter   & Typical            & Photon Count      		& Height of \\
                 &  Wavelength  &  Bandpass & Exp Time      & Statistics          		& Formation \\
	         &  (\AA)       &  (\AA)    & (ms)               & (pixel$^{-1}$s$^{-1}$)   	& (km)	\\
\hline
Ca-K core               &   3933.7     		&    1.00        & 200         & 68,000         &  $< 1300^{a}$ \\
Blue continuum          &   4170.0     		&    52.00       & 10          & 1,928,000      &  $< 250$ \\
Blue continuum		&   3501.0     		&    102.00	 & TBD         & TBD		&  $< 250$     \\
G-band			&   4305.5     		&    9.20	 & 15          & 1,189,000	&  $< 250^{b,c}$ \\
Magnetograms            &   6302.5     		&    $0.21^{d}$  & 240         & 80,000    	&  $< 180^{e}$ \\
UBF			&   variable$^{d}$      &    $0.21^{d}$  & variable    & variable    	&  variable \\
H$\alpha$ core (Zeiss)	&   6562.8     		&    $0.25^{f}$  & 240         & 42,000		&  $< 1500^{g}$ \\
\hline
\end{tabular}
\footnotesize \\
$^{a}$: \opencite{Bee69} \\
$^{b}$: \opencite{Uit06} \\
$^{c}$: \opencite{Rim04} \\
$^{d}$: \opencite{Bec75} \\
$^{e}$: \opencite{Ala90} \\
$^{f}$: \opencite{Bal02} \\
$^{g}$: \opencite{Ver81} \\
\end{table}

\section{Observing Modes}
\label{observing} 

ROSA was successfully commissioned on the 76~cm DST during August 2008 (Fig~\ref{ROSA_bench}). It utilizes the 
DST's large optical benches and high-order adaptive optics to achieve multi-wavelength observations 
of the lower solar atmosphere. Due to a wide range of optical components available, a large number of filter combinations 
can be achieved. However, details of several key optical elements are 
outlined here. 


A Zeiss universal birefringent filter (UBF; \opencite{Bon89}) may be implemented for narrowband 
($\approx$0.2~\AA) imaging. However, due to a significant decrease in the transmission profile 
of the UBF at shorter wavelengths ($\sim 0.4$\% at 4000~\AA\, compared to $\sim 7.2$\% at 
7000~\AA; \opencite{Bec75}), it is more desirable to allow the UBF to obtain filtergrams in 
the ``red'' portion of the optical spectrum. Typically, the UBF is most commonly used for 
chromospheric H$\alpha$ core imaging at 6562.8~\AA. However, in addition to optical imaging, 
the UBF may be used in conjunction with a Wollaston prism to capture photospheric magnetic field 
information at high spatial and temporal resolution. By tuning the UBF into the wing of the 
magnetically sensitive Fe~{\sc{i}} (6302.5~\AA) absorption line, two ROSA cameras may be used to 
capture simultaneous Stokes~V observations of left- and right-hand circularly polarized light, 
allowing the magnitude of the line-of-sight magnetic field component to be studied. If it is 
desirable to obtain magnetic-field information in the photosphere, simultaneous narrowband H$\alpha$ core 
imaging can still be obtained using an additional tunable Zeiss filter.

For imaging in the ``blue'', independent Halle/Oriel filters may be used 
which have a more efficient transmission curve for this portion of the electromagnetic 
spectrum. Examples include wideband continuum and G-band filters for photospheric imaging, 
as well as a Calcium-K core filter for observations of the 
upper-photosphere/lower-chromosphere. Typically, observers will choose to reduce the 
overall field-of-view size in exchange for higher spatial resolution. In order to do this, 
acquired images should have a sampling of $\approx$0.07$''$ per pixel to match the telescope's 
diffraction-limited resolution in the ``blue'' portion ({\eg} $4170$~\AA) of the optical 
spectrum to that of the CCD. Doing so places $50$~km of the solar surface on each pixel, 
providing $100$~km spatial resolution and an overall field-of-view size of $50200 \times 50100$~km$^{2}$. 
Maintaining the same spatial sampling across all ROSA cameras will insure the resulting 
data has the same field-of-view. However, users may choose to match the telescope's 
diffraction-limited resolution in each of their desired bandpasses to that of the CCD, thus 
providing optimal spatial resolution on all ROSA cameras.

Details of the most common filters may be found in Table~\ref{table2}. 
Exposure times listed are typical of quiet-Sun, disk-centre observations filling approximately one-third of the well depth 
per exposure. Utilizing count rates obtained during the ROSA commissioning run (incorporating the default plate-scale sampling 
of 0.07$''$ per pixel) and the camera's quantum efficiency curve (upper panel of Figure~\ref{QE_Noise}), photon 
count-rate statistics can be derived, as listed in the fifth column of Table~\ref{table2}.
Since ROSA utilizes a host of independent computers and storage devices, it is not necessary 
to run all six cameras during a single observing sequence. Indeed, any combination of 
cameras may be used depending on the observers scientific requirements.

\section{User Interface}
\label{interface}

\begin{figure} 
\centerline{\includegraphics[width=1.0\textwidth,clip=]{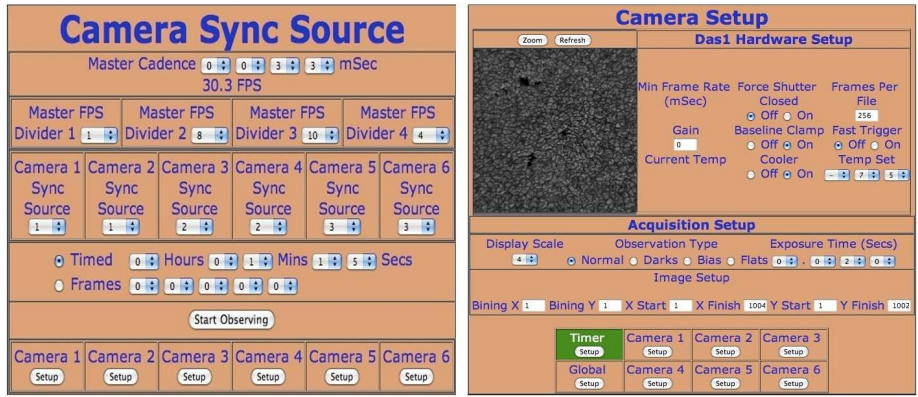}}
\caption{Screenshots of the ROSA GUI during system synchronization (left) and camera setup (right). Key input parameters 
can be set such as the master and secondary frame rates, which cameras are active and the desired system run time (left) in addition 
to the exposure time, shutter setting and camera temperature (right) before an observing sequence has commenced.}
\label{GUI}
\end{figure}

A user can individually adjust camera operating temperatures, frame rates and exposure 
times through use of a simple graphical user interface (GUI; Fig~\ref{GUI}). The GUI utilizes a 
Java-script interface to create a stable 
platform incorporating drop-down menus in an easy-to-follow environment. 
Through the GUI, users can also 
select which cameras to run, and modify their triggering sequence to maximize 
individual camera frame rates and insure the synchronization of specific cameras. 

The GUI also allows the user to select which type of observations will be 
acquired by selecting the desired option. These options include solar data acquisition, 
dark images with a closed shutter, bias frames using a zero-second exposure 
time and flat field images. Each choice will result in the saved filename 
being modified to include the desired option, thus making post-acquisition 
file searching much easier. Once an observing sequence has commenced, the GUI 
displays sample images coming from each of the cameras, refreshed approximately 
every second. A detailed description of the GUI and a guide to 
its adjustable parameters may be found in the ROSA User Manual, available at 
{\ttfamily{http://star.pst.qub.ac.uk/rosa}}.

\newpage
\section{Calibration Data}
\label{calibration}

Due to the high sensitivity of the ROSA system, it is imperative to accompany 
data imaging sequences with a range of additional observations. It is these 
supplementary observations which allow accurately processed images to be produced 
via the ROSA-specific data reduction pipeline. To compensate for dark current 
and read-out noise, sufficient dark images ($>250$) with exposure times equal 
to the data acquisition must be obtained and subtracted from the science 
images. Many flat-field images ($>500$) should also be taken through an un-flat 
telescope mirror coupled to a random guider to allow an accurate gain table to 
be created for each ROSA camera. These key calibration procedures will enable 
science data to be corrected for camera inconsistencies as well as for variable 
light levels across the incident beam. At this stage, PFIR 
techniques such as speckle \cite{Wei83} or 
multi-object multi-frame blind deconvolution \cite{van05} may be implemented. Indeed, 
\inlinecite{Wog08} have shown how modern PFIR techniques can produce reconstructed images 
which are photometrically accurate, allowing reliable studies of solar structures to 
be undertaken.



However, image reconstruction processes are 
extremely CPU intensive and this must be taken into consideration before attempting to 
reduce the data. For example, during the commissioning run a single ROSA camera 
obtained over 105,000 raw G-band filtergrams in only one hour of observing. 
With excellent seeing 
conditions, an image reconstruction of $16 \rightarrow 1$ was deemed suitable, providing 
a reconstructed cadence of 0.5~s. Thus, over 6500 separate reconstructions were 
required, with each individual process taking approximately 30~CPU~minutes to 
complete on a modern Intel Xeon processor. If access to only one computer is possible 
({\eg} preparing the data solely on the user's desktop PC), then reconstructing 
all 6500 images would take in excess of 135~days, with additional time being required 
to process the data from the remaining five ROSA cameras. As telescope schedules often 
grant individuals in excess of seven days observing time, and each day commonly 
provides $2 - 3$~hours of good seeing conditions, it is imperative to consider how 
and where the acquired observations will be reduced so excessive time delays are not 
experienced.

\begin{figure} 
\centerline{\includegraphics[width=1.0\textwidth,clip=]{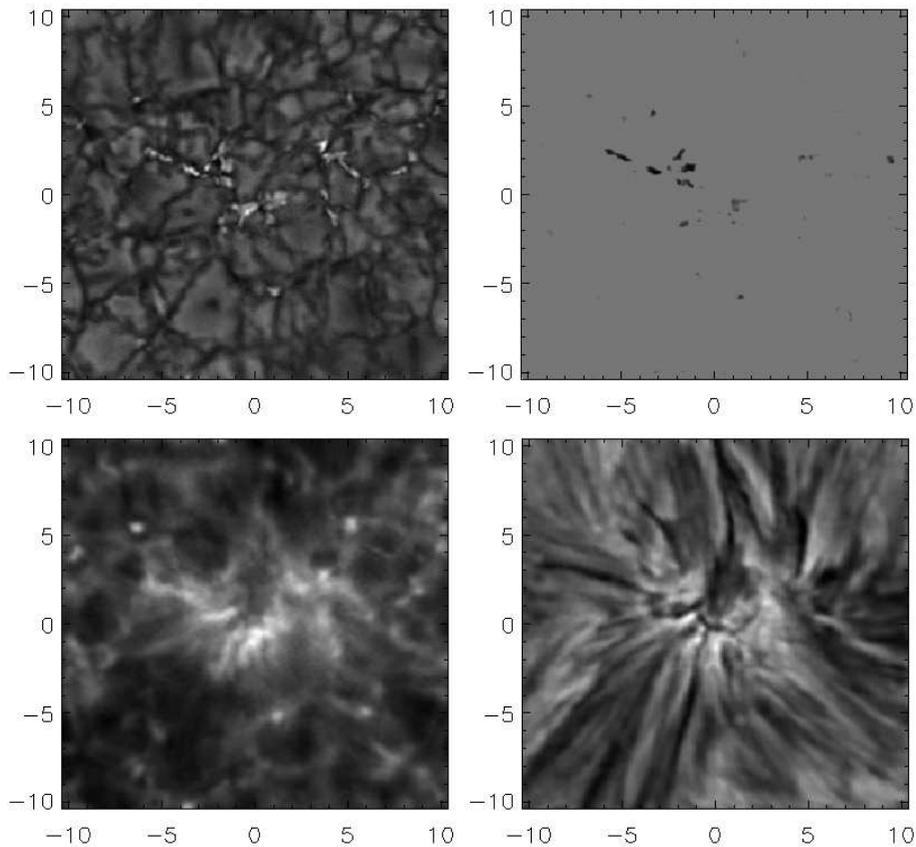}}
\caption{Simultaneous images acquired with ROSA during the commissioning run in August 2008. The top left panel 
shows a collection of magnetic bright points visible as intensity enhancements through the G-band filter, whereas 
the top right panel reveals a line-of-sight magnetogram established from 
difference imaging of the Stokes~V parameters obtained from the magnetically sensitive Fe~{\sc{i}} absorption 
line at 6302.5~\AA. The lower left image is a co-temporal and co-spatial representation of the 
upper-photosphere/lower-chromosphere through the Ca-K core filter and the lower right panel is how the collection 
of magnetic bright points look through the chromospheric H$\alpha$ filter. The field-of-view (axes in arcseconds) 
shown here is approximately $20'' \times 20''$ or $15000 \times 15000$~km$^{2}$.}
\label{ROSA_science}
\end{figure}

In order to co-align each ROSA camera, additional calibration images should 
be acquired. Images of an Air-Force target will allow compensation for general rotation and image 
mirroring, while acquiring a burst of grid images will enable 
small-scale inter-camera image rotation to be evaluated and removed. With these 
camera positioning artifacts removed, image destretching can be accomplished through 
use of the ROSA data reduction pipeline. Under normal circumstances, a 
$40 \times 40$ grid, equating to a $1.7''$ separation between spatial samples 
(for diffraction-limited resolution at 4170~{\AA}), is used to evaluate local 
offsets between successive images, allowing compensation for spatial distortions 
caused by atmospheric turbulence and/or air bubbles crossing the entrance aperture 
of the telescope. The fine destretching grid implemented in this process allows 
compensation for small-scale seeing conditions of $1''$ to $2''$ in size.

To help facilitate and expedite data preparation, QUB have developed a parallel processing 
cluster dedicated to ROSA data reduction. This cluster consists of 25 Intel Xeon quad-core CPUs providing 100 
processing nodes on a dedicated private network. Users of the ROSA instrument may opt for 
all data reduction to be carried out on the QUB cluster, or may avail of the freely distributed ROSA 
pipeline for use on their own reduction cluster.
The accurate reconstruction, co-alignment and destretching of multiple atmospheric heights will promote 
the use of multi-wavelength studies, whereby the cause of dynamic phenomena can be probed 
and analysed, with the subsequent effect on other layers of the solar atmosphere evaluated.

\section{Conclusions} 
\label{conclusions}

In recent years, the solar community has expressed a strong desire for a high-cadence, 
synchronized, multi-camera system to be introduced which would allow unprecedented 
spatial {\it{and}} temporal resolutions of solar structures over a wide range of atmospheric 
heights. With the successful commissioning of the ROSA instrument, it is now possible to 
observe solar structures at high cadences in up to six independent wavelengths simultaneously. 
Figure~\ref{ROSA_science} shows the capabilities of observing in at least four synchronized 
wavelengths (G-band filtergrams, magnetograms from circularly-polarized light at 6302.3~\AA, 
Ca-K core imaging at 3933.7~{\AA} and narrowband H$\alpha$ core filtergrams). These types of high-resolution 
images present an excellent opportunity to scientists, whereby magnetic structures can be 
investigated at multiple atmospheric heights with temporal resolutions exceeding ten frames per second.

Due to the rack-mounted nature of the ROSA instrument, future upgrades to the system 
can be readily carried out. As technology improves and the size of components decrease, 
ROSA hardware can be updated continually to keep it at the forefront of high-cadence 
imaging. Such upgrades will include the increased capacity of onboard storage, an improved 
connectivity for external media ({\eg} e-SATA and firewire connections) and a turbulence-reducing 
method of camera cooling ({\eg} via water cooling). 
With the presence of backside illuminated CCDs now appearing in the 
marketplace, existing ROSA cameras may be replaced with such devices. In instances where low 
exposure times are required, this new breed of camera may lead to performance boosts in 
photon-starved regimes by offering a higher quantum efficiency when compared to traditional front-illuminated devices. 
The ability to adapt ROSA with continually 
changing demands and computer architecture has identified ROSA as a bridging instrument 
for the Advanced Technology Solar Telescope and places a high-cadence solar imager in line 
to become a first-light instrument on this revolutionary facility in 2015.

%

%

%

%
\begin{acks}
This work was supported by the U.K. Science and Technology Facilities Council and DBJ wishes to thank the Science and 
Technology Facilities Council for the award of a Post-Doctoral Fellowship. FPK is grateful to AWE Aldermaston 
for the award of a William Penney Fellowship. PJC thanks the Northern Ireland Department for 
Employment and Learning for a PhD studentship. Observations were obtained at the National Solar Observatory, operated by 
the Association of Universities for Research in Astronomy, Inc. (AURA), under cooperative agreement with the National Science 
Foundation. Finally we would like to thank the technical staff at the DST for perseverance in the face of 
seemingly endless camera and filter arrangements.
\end{acks}

%
%
%

\end{article} 
\end{document}